\documentclass[twocolumn]{aastex631}

\newcommand{\ariii}{[\ion{Ar}{3}]}
\newcommand{\arii}{[\ion{Ar}{2}]}
\newcommand{\neiii}{[\ion{Ne}{3}]}
\newcommand{\neii}{[\ion{Ne}{2}]}
\newcommand{\nev}{[\ion{Ne}{5}]}
\newcommand{\siii}{[\ion{S}{3}]}
\newcommand{\siv}{[\ion{S}{4}]}
\newcommand{\oii}{[\ion{O}{2}]}
\newcommand{\oiii}{[\ion{O}{3}]}
\newcommand{\oiv}{[\ion{O}{4}]}
\newcommand{\feii}{[\ion{Fe}{2}]}
\newcommand{\hm}{${\rm H_2}$}
\newcommand{\mm}{${\rm \mu m}$}
\definecolor{lightblue}{rgb}{0., 0.398, 0.796}
\definecolor{red}{rgb}{1, 0, 0}

\usepackage{lipsum}
\usepackage{multirow}
\usepackage{gensymb}
\usepackage{hyperref}
\usepackage{amsmath}	
\usepackage{amssymb}	

\usepackage[dvipsnames]{xcolor}

\usepackage{booktabs}

\begin{document}

\title{JWST/MIRI-MRS view of the metal-poor galaxy CGCG 007-025: the spatial location of PAHs and very highly ionized gas}

\shortauthors{del Valle-Espinosa et al.}
\shorttitle{PAH and \nev~emission in Metal Poor Galaxies}

\author[0000-0002-0191-4897]{Macarena G. del Valle-Espinosa}
\affiliation{Space Telescope Science Institute, 3700 San Martin Drive, Baltimore, MD 21218, USA}

\author[0000-0003-2589-762X]{Matilde Mingozzi}
\affiliation{AURA for ESA, Space Telescope Science Institute, 3700 San Martin Drive, Baltimore, MD 21218, USA}

\author[0000-0003-4372-2006]{Bethan James}
\affiliation{AURA for ESA, Space Telescope Science Institute, 3700 San Martin Drive, Baltimore, MD 21218, USA}

\author[0000-0003-4945-0056]{Rub\'{e}n S\'{a}nchez-Janssen}
\affiliation{Isaac Newton Group, Tenerife, Canary Islands, Spain}

\author[0000-0001-9490-899X]{Juan Antonio Fern\'{a}ndez-Ontiveros}
\affiliation{Centro de Estudios de F\'isica del Cosmos de Arag\'on (CEFCA), Plaza San Juan 1, E--44001 Teruel, Spain}

\author[0000-0001-9719-4080]{Ryan J. Vaught}
\affiliation{Space Telescope Science Institute, 3700 San Martin Drive, Baltimore, MD 21218, USA}

\author[0000-0001-5758-1000]{Ricardo O. Amor\'{i}n}
\affiliation{Instituto de Astrof\'{i}sica de Andaluc\'{i}a (CSIC), Apartado 3004, 18080 Granada, Spain}

\author[0000-0001-9162-2371]{Leslie Hunt}
\affiliation{INAF - Osservatorio Astrofisico di Arcetri, Largo E. Fermi 5, I-50125, Firenze, Italy}

\author[0000-0003-4137-882X]{Alessandra Aloisi}
\affiliation{Space Telescope Science Institute, 3700 San Martin Drive, Baltimore, MD 21218, USA}
\affiliation{Astrophysics Division, Science Mission Directorate, NASA Headquarters, 300 E Street SW, Washington, DC 20546, USA}

\author[0000-0002-2644-3518]{Karla Z. Arellano-C\'{o}rdova}
\affiliation{Institute for Astronomy, University of Edinburgh, Royal Observatory, Edinburgh EH9 3HJ, UK}

\author[0000-0002-4153-053X]{Danielle A. Berg}
\affiliation{Department of Astronomy, The University of Texas at Austin, 2515 Speedway, Stop C1400, Austin, TX 78712, USA}

\author[0000-0002-0302-2577]{John Chisholm}
\affiliation{Department of Astronomy, The University of Texas at Austin, 2515 Speedway, Stop C1400, Austin, TX 78712, USA}

\author[0000-0001-8587-218X]{Matthew Hayes}
\affiliation{Stockholm University, Department of Astronomy and Oskar Klein Centre for Cosmoparticle Physics, AlbaNova University Centre, SE-10691, Stockholm, Sweden}

\author[0000-0003-4857-8699]{Svea Hernandez}
\affiliation{AURA for ESA, Space Telescope Science Institute, 3700 San Martin Drive, Baltimore, MD 21218, USA}

\author[0000-0002-2954-8622]{Alec S.\ Hirschauer}
\affiliation{Department of Physics \& Engineering Physics, Morgan State University, 1700 East Cold Spring Lane, Baltimore, MD 21251, USA}

\author[0000-0002-1706-7370]{Logan Jones}
\affiliation{Space Telescope Science Institute, 3700 San Martin Drive, Baltimore, MD 21218, USA}

\author[0000-0001-9189-7818]{Crystal L. Martin}
\affiliation{Department of Physics, University of California, Santa Barbara, Santa Barbara, CA 93106, USA}

\author[0000-0002-3258-3672]{Livia Vallini}
\affiliation{INAF – Osservatorio di Astrofisica e Scienza dello Spazio di Bologna, Via Piero Gobetti, 93/3, I-40129 Bologna, Italy}

\author[0000-0002-9217-7051]{Xinfeng Xu}
\affiliation{Department of Physics and Astronomy, Northwestern University, 2145 Sheridan Road, Evanston, IL 60208, USA}
\affiliation{Center for Interdisciplinary Exploration and Research in Astrophysics (CIERA), Northwestern University, 1800 Sherman Avenue, Evanston, IL 60201, USA}

\correspondingauthor{Macarena G. del Valle-Espinosa}
\email{mgarciavalle@stsci.edu}

\begin{abstract}

Polycyclic Aromatic Hydrocarbons (PAHs) are key diagnostics of the physical conditions in the interstellar medium and are widely used to trace star formation in the mid-infrared (mid-IR). The relative strengths of mid-IR PAH features (e.g., 6.2, 7.7, 11.3~$\mu$m) are sensitive to both the size and ionization state of the molecules and can be strongly influenced by the local radiation field. However, at low metallicities (Z$<$0.2Z$_\odot$), detecting PAHs remains notoriously difficult, likely reflecting a combination of suppressed formation and enhanced destruction mechanisms. We present new JWST/MIRI MRS observations of the metal-poor (Z$\sim$0.1Z$_\odot$) dwarf galaxy CGCG~007-025. We confirm the tentative PAH detection previously reported from Spitzer data and, for the first time, identify a compact ($\sim50$~pc) PAH-emitting region nearly co-spatial with the newly detected \nev~(I.P.~$\sim97$eV) emission and the galaxy’s most metal-poor, strongly star-forming region. The PAH$_{11.3\mu m}$ and PAH$_{12.7\mu m}$ features are detected, while no emission is found from other typically brighter features, suggesting a PAH population dominated by large, neutral molecules resilient to hard ionizing fields. When compared with models, mid-IR line ratios involving \neiii, \oiv, and \nev\ can only be reproduced by a combination of star formation and AGN ionization, with the latter contributing 4--8\%. The \oiv\ and \nev\ luminosities exceed what massive stars or shocks can produce, highlighting a puzzling scenario in line with recent JWST observations of similar galaxies. This work provides a crucial reference for studying the physical conditions of the dust and star formation in low-metallicity starburst regions, environments typical of the early universe.

\end{abstract}



\section{Introduction} \label{sec:intro}

Polycyclic Aromatic Hydrocarbons (PAHs) are widespread components of the interstellar medium (ISM), producing strong vibrational features in the near- and mid-infrared (IR) at 3.3, 6.2, 7.7, and 11.3 \mm~\citep[see][for a review]{li2020NatAs...4..339L}. These features often dominate the mid-IR spectra of galaxies and serve as key tracers of the physical conditions in the ISM and star formation intensity \citep[][among others]{sales2010ApJ...725..605S,shipley2016ApJ...818...60S}. More importantly, PAHs have been used as tracers of the dust content and composition in galaxies \citep[e.g.,][]{haas2002A&A...385L..23H,jones2015MNRAS.448..168J,sutter2024ApJ...971..178S}. Understanding their role is particularly crucial in low-metallicity environments \citep[${\rm12 + log(O/H) < 8.0~or~Z< 0.2Z_\odot}$,][]{wu2006ApJ...639..157W,hunt2010ApJ...712..164H}, where the origin and survival of dust remain poorly constrained. A recent JWST study by \citet{shivaei2024A&A...690A..89S}
has found that the PAH mass fraction in galaxies at $z = 0.7$–2 declines sharply below ${\rm log(M_*/M_\odot) \lesssim 9.7}$ and ${\rm 12 + log(O/H) \lesssim 8.4~(Z \lesssim0.5Z_\odot)}$, consistent with a combination of enhanced PAH destruction in strong radiation fields and reduced efficiency of PAH formation at low metallicities. This picture has been challenged by recent observations of high-redshift galaxies. Since the metallicity of the overall galaxy population decreases with redshift \citep[e.g.,][]{curti2023MNRAS.518..425C}, one might expect low dust content in the early Universe. However, observations of unexpectedly dusty galaxies at very high redshift (up to $z\sim8$) indicate that substantial dust reservoirs were already in place when asymptotic giant branch stars could not yet have contributed significantly to the dust budget \citep{witstok2023Natur.621..267W,omerod2025MNRAS.tmp.1180O}, suggesting that both the production and destruction channels of dust and complex molecules such as PAHs are not yet fully understood.

In this letter, we focus on the nearby \citep[${z \sim 0.004875}$; ${\rm d_L = 23 \pm 5 ~Mpc}$,][]{kourkchi2020ApJ...902..145K}, metal deficient \citep[$\sim 10$\%~Z$_\odot$ or 12 + log(O/H) $\sim$7.7,][among others]{izotov2004ApJ...602..200I, zee2006ApJ...636..214V, izotov2007ApJ...662...15I} dwarf galaxy \citep[${\rm M_{\star} \sim 1.2 \times 10^8 ~M_\odot}$,][]{marasco2023A&A...670A..92M} CGCG 007-025. This starbursting galaxy displays a prominent star-forming region with ${\rm log~(sSFR [yr^{-1}]) \approx -7.61}$ and \oiii$\lambda$5007/\oii$\lambda\lambda$3727 $\approx$ 2.9, satisfying the criteria to be treated as a local analogue of high-redshift galaxies \citep[e.g.,][]{berg2022ApJS..261...31B,brammer2012ApJ...758L..17B}. Previous observations of this galaxy with Spitzer/IRS revealed the presence of very weak PAH features located at 7.7 and 11.3 \mm~\citep{hunt2010ApJ...712..164H}. Using new JWST MIRI/MRS data, we 
uniquely characterize the properties of PAH emission features in CGCG~007-025 to provide new insights into the survivability of these molecules in low-metallicity environments, similar to those of the first galaxies in the Universe.

In Section \ref{sec:data} we describe the MIRI/MRS observations and data reduction, and we present our emission line measurements and PAH modeling methodology in Section \ref{sec:methods}. The main results on the spatial extension of the PAH features, together with the general properties of the emitting region, are compiled in Section \ref{sec:results}. The discussion is in Section \ref{sec:discussion}. We summarize our findings in Section \ref{sec:conclusions}.

%
%

\section{Data} \label{sec:data}

\begin{figure}
    \centering
    \includegraphics[width=0.45\textwidth, clip, trim=0 0 350 0]{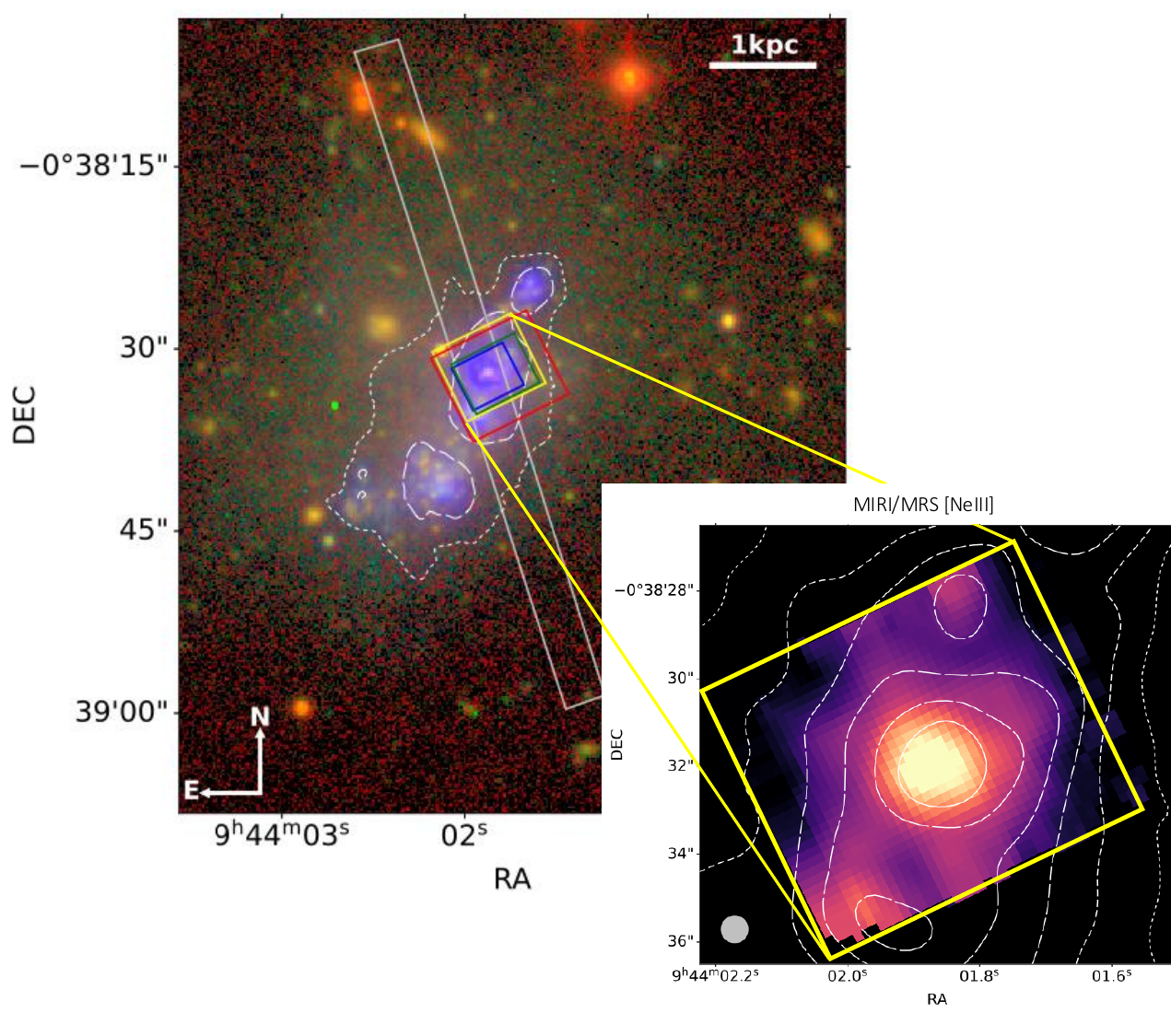}
    \caption{SUBARU/HSC $giy$ color image of the high-$z$ local analog CGCG 007-025, where \textit{g, i} and \textit{y} filters correspond to the colors blue, green and red, respectively. Overlaid are the MIRI/MRS channel coverages displayed in blue (channel 1), green (channel 2), yellow (channel 3) and red (channel 4). The slit position from the previous Spitzer/IRS observations is shown in light gray. \textit{Inset panel}: \neiii\,$\lambda$15.55${\rm\mu m}$ integrated flux (channel 3, PSF FWHM represented in the bottom left of the inset), showing a PSF-like structure as well as  more extended diffuse emission. In both images, the white contours trace the H$\alpha$ emission as seen by MUSE \citep[from][]{valleespinosa2023MNRAS.522.2089D}, with the outermost corresponding to H$\alpha$ emission down to $\sim 1\times 10^{-16}$ erg/s/cm$^2$.}
    \label{fig:cgcgpresentation}
\end{figure}

This letter presents the JWST MIRI/MRS data of CGCG~007-025. The main starbursting region of the galaxy was observed as part of the Cycle~2 JWST program  GO~4278 (PI: Mingozzi). Figure \ref{fig:cgcgpresentation} shows the MIRI/MRS field of view (FoV) coverage superimposed on the Subaru/HSC $giy$ color image. The data were reduced using the JWST data reduction pipeline version~\textit{1.17.1.dev} with the CRDS~1338 context (see Appendix \ref{sec:appendixData} for more details). The final data products were the 12 sub-channel cubes with the default spatial and spectral sampling, whose astrometry matches with HST/WFC3 F814W filter. The data show a PSF-like structure (see inset of Figure \ref{fig:cgcgpresentation}), which causes fringe patterns to vary systematically across the PSF and requires further corrections \citep[see][for a full description of the mitigation process we applied]{mingozzi2025ApJ...985..253M}.

%
%

\section{Methodology} \label{sec:methods}

Depending on the signal-to-noise ratio (S/N) of the continuum, we adopted two complementary approaches to analyze the mid-IR spectral features of CGCG~007-025. For the integrated spectrum, where the S/N is high (S/N~$ > 20$ in the continuum), we modeled the continuum and spectral features using \textsc{pahfit} \citep{smith2007ApJ...656..770S}. At the spaxel level, where the continuum S/N is lower, we instead performed localized fits in narrow spectral windows around each feature of interest.

\subsection{Extended ionized gas and PAH emission} \label{sec:methext}

To characterize the extended emission of the ionized gas at the spaxel level, we used an approach similar to the one described in \citet{mingozzi2025ApJ...985..253M}. We used a combination of a Gaussian profile and a 1-degree polynomial --to account for emission-line and surrounding continuum emission-- within a $\pm10^3$ km/s window around each line of interest. We fit the emission lines listed in Table \ref{tab:fluxes}.

To characterize the extension of the PAH emission, we focus our analysis on the 11.3 \mm~bump --the brightest PAH feature present in this object (see Section \ref{sec:resultsint}). We created a sliced cube (from the original channel 2 cube) of $\pm10^4$ km/s around this feature and applied a moving median smooth filter over 71 spectral pixels to reduce the noise on the bump \citep[similarly to][]{hunt2025ApJ...993...84H}, since this feature appears especially weak when using individual spaxels. After smoothing, we modeled the spectra using a combination of a 1-degree polynomial and a Drude profile \citep[the standardized profile for PAH emission, see][]{draine2007ApJ...657..810D}.

\begin{figure*}
    \centering
    \includegraphics[trim = 0 0 550 0,clip,width=0.35\textwidth]{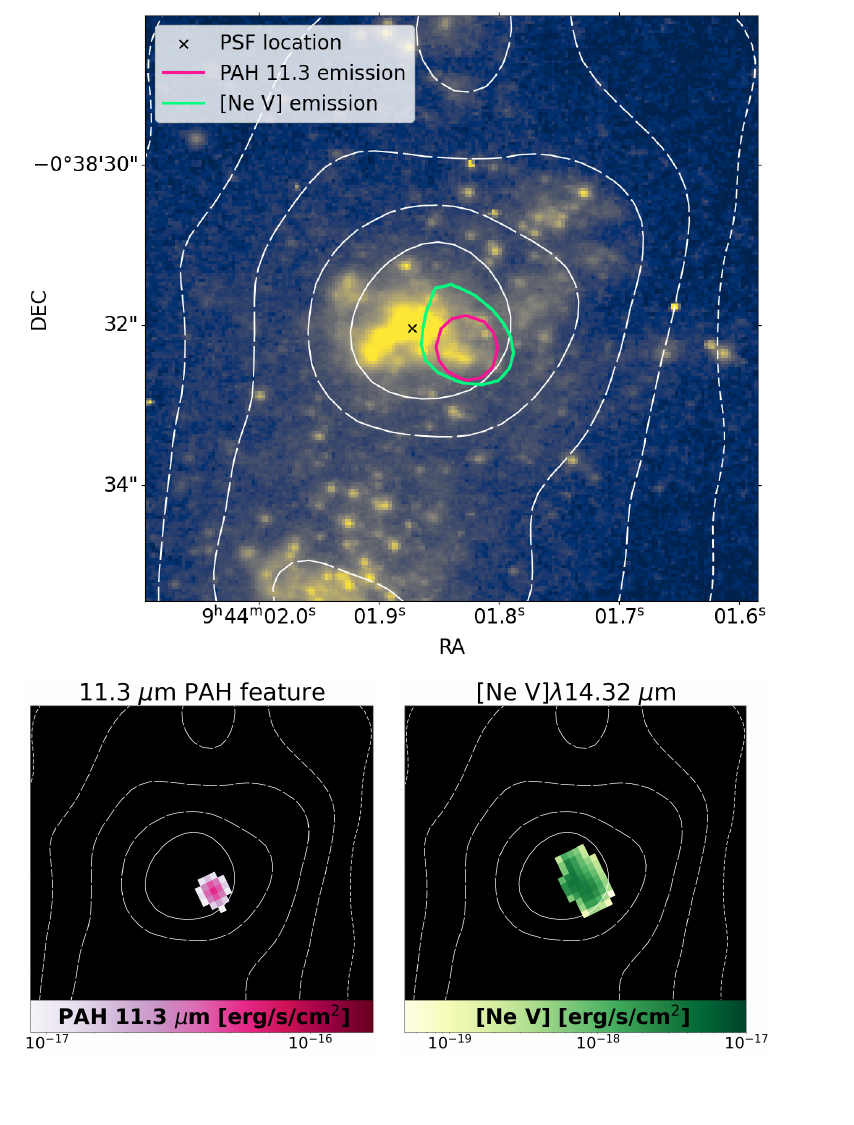}
    \includegraphics[width=0.6\textwidth]{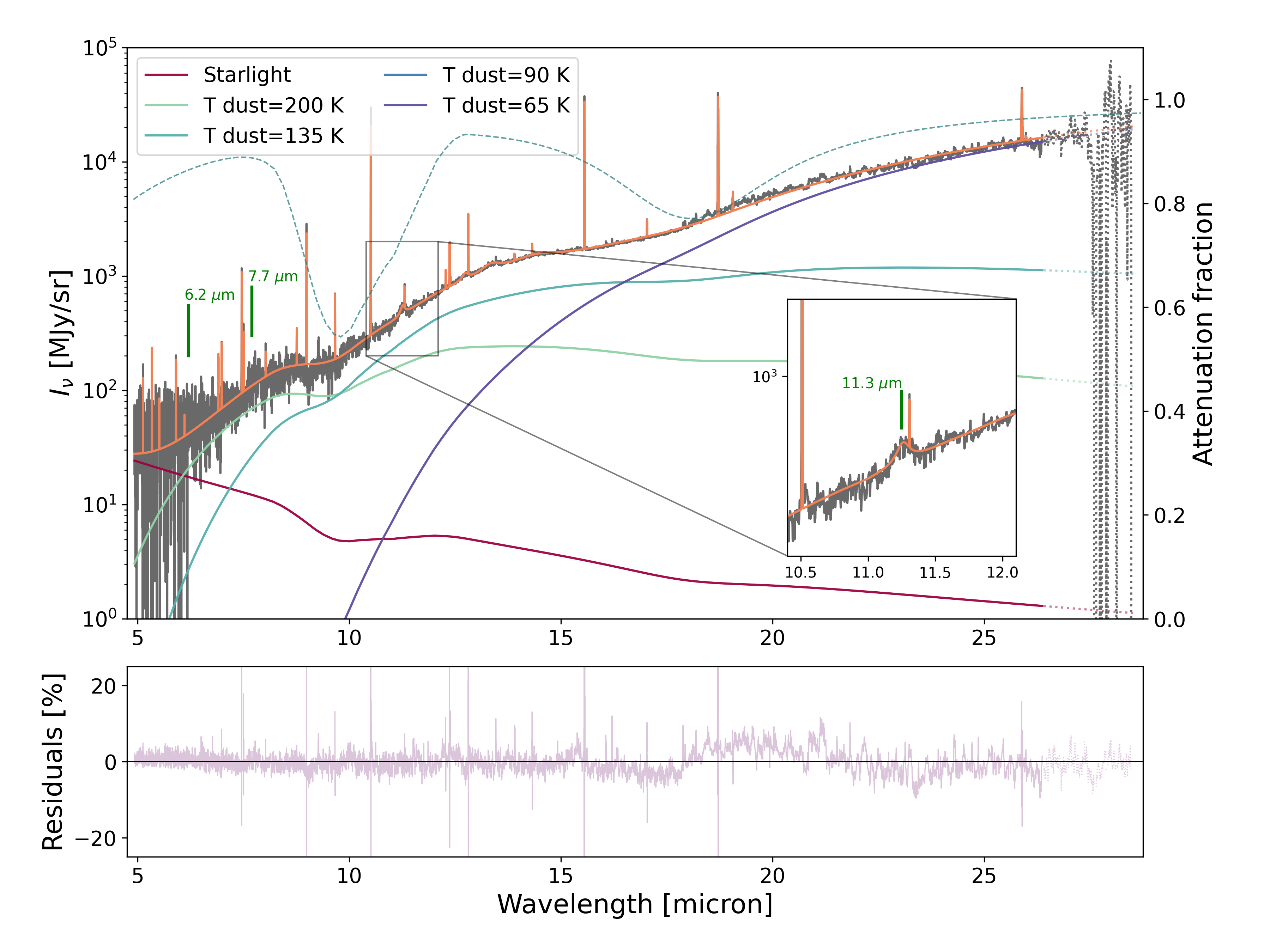}
    \caption{\textit{Top left panel}: Location of the 11.3 \mm~PAH feature. The solid pink contour encompasses the region were the 11.3 \mm~PAH feature is detected above $3\sigma$ on the smooth spectra. The \nev~extended emission, detected at spaxel level with ${\rm S/N > 3}$, is delimited by the solid light green contour. The location of the PSF-like structure is marked with a black cross. The background image corresponds to the F814W HST/WFC3 filter, with the white contours tracing the H$\alpha$ emission as seen by MUSE  \citep[from][]{valleespinosa2023MNRAS.522.2089D}. 
    \textit{Bottom left panels}: Flux distribution of the 11.3 \mm~PAH feature (pink distribution, left) and \nev~extended emission (green distribution, right) from which the contours displayed on the top panel have been obtained. \textit{Right panel}: \textsc{pahfit} modeling of the PAH emitting region in CGCG 007-025. The spectrum is shown in solid-gray, with the best fit modeling displayed in solid-orange and the residuals in light purple in the bottom panel. The dotted-lines show parts of the spectrum not used during the fit. The derived values for the attenuation curve (dashed line) are represented in the right hand side axis of the top plot. In the axis inset, we zoom around the 11.3 \mm~PAH feature. The theoretical location of other PAH features such as the 6.2 and the 7.7 \mm, not present in the data, are highlighted with vertical green lines.}
    \label{fig:pahMapandSpectrum}
\end{figure*}

\subsection{Integrated emission} \label{sec:methint}

Prior to extraction, we matched the spatial resolution across bands using the linear relation between the MIRI/MRS PSF FWHM and wavelength from \citet{law2023AJ....166...45L} as a first approximation for the PSF variation between channels. At each wavelength step, we convolved the data with a 2D Gaussian kernel whose width matches the difference in quadrature between the PSF FWHM at that wavelength and the reference value. We considered the reference value to be that of [\ion{O}{4}]\,$\lambda$25.89, corresponding to a ${\rm FWHM_{PSF} \sim 3}$px (0\farcs95, 98 pc). After convolution, we resampled the cubes to match the pixel scale of channel 1 (0\farcs13). 

Once the integrated spectrum is on hand, we use \textsc{pahfit} to model the mid-IR continuum of CGCG 007-025. \textsc{pahfit} fits simultaneously the dust continuum, PAH features and emission lines using a combination of blackbody, Drude and Gaussian profiles, respectively, together with an extinction curve \citep[][]{smith2007ApJ...656..770S}. We update the default list of emission lines to be modeled by \textsc{pahfit} using the line list presented in \citet{vandeputte2024A&A...687A..86V}. Moreover, we allowed the PAH profiles to vary in central wavelength ($\pm$1\%) and in FWHM (-60\%, +10\%), similarly to \citet{donnan2023MNRAS.519.3691D}. 

%
%

\section{Results} \label{sec:results}

\subsection{Location and extension of the\,11.3\,$\mu m$\,PAH feature} \label{sec:resultsext}

We present, in the left panel of Figure \ref{fig:pahMapandSpectrum}, the spatially resolved 11.3~\mm~PAH feature. The 11.3 \mm~PAH emission peaks near the location of one of the clumps in CGCG 007-025, with an extension of $\sim$ 50 pc (delimited by the pink contour). At this wavelength, the native PSF has a ${\rm FWHM_{PSF}}$ = 2.8 px (0\farcs48, 48 pc), so that the region is barely spatially resolved in channel 2. We also show, in pink, the flux distribution of this feature in the bottom left panel of Figure \ref{fig:pahMapandSpectrum}. We used this contour to extract a spectrum which maximizes the signal of the 11.3 \mm~feature (see right panel of Figure \ref{fig:pahMapandSpectrum}). 

%
%

\subsection{Integrated spectrum of the PAH emitting region} \label{sec:resultsint}

Following the method described in Section \ref{sec:methint}, and using the aperture identified on the PAH map, we extracted the spectrum of the PAH emitting region (solid gray line in the right panel of Figure \ref{fig:pahMapandSpectrum}).
The best fit model from \textsc{pahfit}, together with the residuals, are displayed in the same figure in light orange and light purple, respectively. We also zoom around the 11.3 \mm~PAH to highlight the performance of \textsc{pahfit} in reproducing this feature. In Table \ref{tab:fluxes} (see Appendix~\ref{sec:appendixSFR}) we compile all the line fluxes measured by \textsc{pahfit}. These fluxes have been corrected for dust attenuation by $1/A$, where $A$ is the attenuation fraction from the \textsc{pahfit} model (dashed gray line in right panel of Figure \ref{fig:pahMapandSpectrum}). For the lines used in the upcoming sections, the attenuation-corrected fluxes are increased almost homogeneously by a factor of $\sim1.1-1.15$, meaning that the line ratios between these lines are almost unaffected by the dust attenuation assumed. It is worth noticing that the attenuation derived from \textsc{pahfit} of $\tau_{silicate} = 1.3$ is mainly constrained by the two apparent absorption features at 9 and 18 \mm, simultaneously. Fitting these wavelength ranges separately return a similar value of the attenuation, which excludes the presence of additional emission features, such as the 14 \mm~ alumina oxidates reported in e.g. \citet{hunt2025ApJ...993...84H}.

From the integrated spectrum shown in Figure \ref{fig:pahMapandSpectrum}, there is no clear evidence of any other PAH emission in CGCG 007-025. However, PAH features such as the 6.2, 7.7 or 8.6 \mm~are typically among the brightest ones \citep[e.g.,][]{hunt2010ApJ...712..164H,sandstrom2012ApJ...744...20S,chastenet2019ApJ...876...62C}.
In attempt to search for these other PAH features, we compare the integrated spectrum of CGCG 007-025 with the \ion{H}{2} spectrum template from PDRs4All \citep[][see Figure \ref{fig:appendixpahs} in the Appendix]{chown2024A&A...685A..75C}. While no clear evidence of emission is seen at 6.2 or 7.7 \mm~, the spectrum of CGCG 007-025 displays an excess of emission compatible with the additional presence of another PAH feature at 12.7 \mm. We measure the flux of this feature by fitting a Drude profile to the continuum subtracted integrated spectrum. In order to obtain upper limits on the non-detected features, we follow the procedure described in \citet{lai2025arXiv250904662L} and report the values in Table \ref{tab:fluxes} (see Appendix \ref{sec:appendixPAHs} for more details).

%
%

\subsection{Ionized gas emission in the mid-IR}
The integrated spectrum of the PAH-emitting region presented in Figure \ref{fig:pahMapandSpectrum} reveals the detection of a plethora of nebular emission lines such as \feii\,$\lambda$5.34\mm, \arii\,$\lambda$6.99\mm, \neii\,$\lambda$12.81\mm, \siii\,$\lambda$18.71\mm, \ariii\,$\lambda$8.99\mm, \siv\,$\lambda$10.51\mm, \neiii\,$\lambda$15.55\mm, and \oiv\,$\lambda$25.89\mm, as well as the high ionization line \nev\footnote{Here we focus only on \nev\,$\lambda$14.32\mm, since the accompanying line at 24.32\mm~falls at the edge of ch4-medium and ch4-long, challenging the detection of the feature.}\,$\lambda$14.32\mm. 
Accordingly, we search for the detection of other lines at the spaxel level, including lines of high ionization potential (IP) such as \oiv~and \nev. More concretely, following the methods outlined in Section~\ref{sec:methext}, we focus on \neii, \neiii, \oiv~and \nev~due to the diagnostic power of these lines (see Section \ref{sec:discussion}). While \neii~and \neiii~emission extend across the entire MIRI FoV (see inset in Figure \ref{fig:cgcgpresentation}), both \oiv~and  \nev~emission are in a more confined configuration, co-spatial with the same compact region as the ${\rm PAH_{11.3~\mu m} }$ extended emission (see light green contour and flux distribution in green in Figure \ref{fig:pahMapandSpectrum}).

%
%

\begin{figure*}
    \centering
    \includegraphics[width=0.85\textwidth]{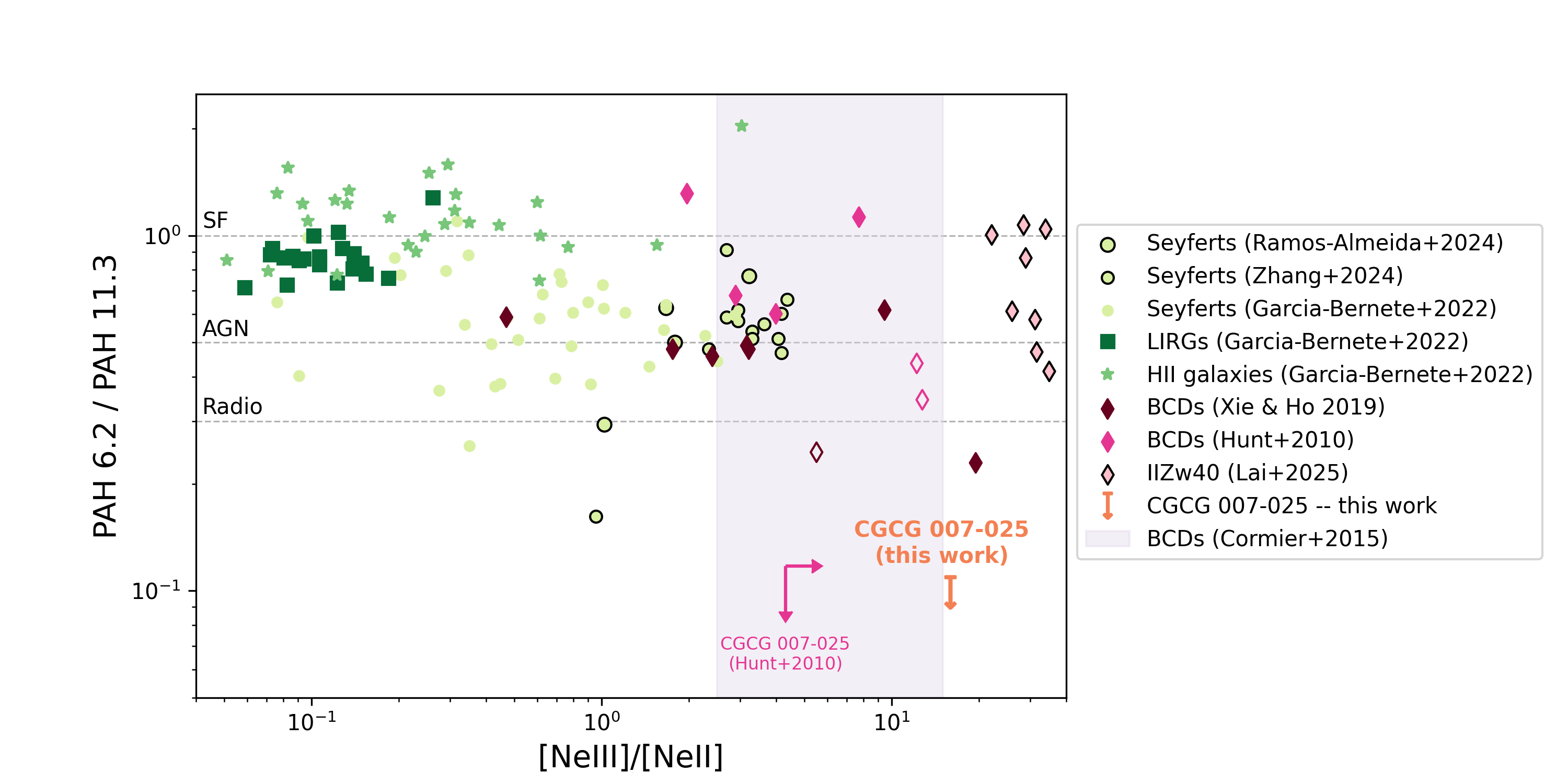}
    \caption{The 6.2 over 11.3 \mm~PAH ratios versus \neiii\,$\lambda$15.55\mm/\neii\,$\lambda$12.81\mm. The location of CGCG 007-025 in the diagram is highlighted with an orange arrow. Comparison samples, using Spitzer/IRS, from the literature are shown for context: star-forming galaxies (light green star), LIRGs (dark green squares), and Seyfert galaxies (lime circles) from \citet{garciabernete2022MNRAS.509.4256G}. Blue compact dwarfs (BCDs) are shown as diamonds (pink for \citet{hunt2010ApJ...712..164H} and purple for \citet{xie2019ApJ...884..136X}), with open symbol indicating upper limits on the PAH ratios. The range of typical \neiii/\neii~ratios for other BCDs is shaded in pink \citep[][]{cormier2015A&A...578A..53C}. Additional Seyfert samples, using MIRI/MRS data, are plotted as lime circles with black edges \citep[from][]{zhang2024ApJ...975L...2Z,ramosalmeida2025A&A...698A.194R}. The recent results from \citet{lai2025arXiv250904662L} on the BCD IIZw40 using MIRI/MRS are shown as light pink diamonds with black edges. The horizontal dashed lines represent the typical PAH ratios for star-forming regions, AGNs and radio galaxies (see text for references).}
    \label{fig:PAH_Nes}
\end{figure*}

\section{Discussion}   \label{sec:discussion}

\subsection{What PAH and line ratios reveal about CGCG\,007-025}

PAH features at different wavelengths can be used to trace the predominant ionization state of the overall PAH population. The emission features at 6.2 \mm~and 7.7 \mm~are typically associated with ionized PAHs, while the feature at 11.3 \mm~is emitted by large, neutral PAHs. 
Consequently, the ratios between PAH features provide insight into the dominant PAH population and the ionizing mechanisms within a galaxy. For instance, several studies have reported thresholds in the 7.7/11.3 and 6.2/11.3 ratios\footnote{In this manuscript, we avoid the use of the 7.7$\mu m$ feature due to the large variations in the continuum noise around this feature (see Figure \ref{fig:appendixpahs}).}, with SF-dominated regions typically showing 6.2/11.3 larger than 1, while environments exposed to harder radiation fields (e.g., AGN) display ratios below 0.3 \citep[e.g.,][]{smith2007ApJ...656..770S,ogle2010ApJ...724.1193O,garciabernete2022MNRAS.509.4256G,donnan2023MNRAS.519.3691D}.

\begin{figure}
    \centering
    \includegraphics[width=0.5\textwidth]{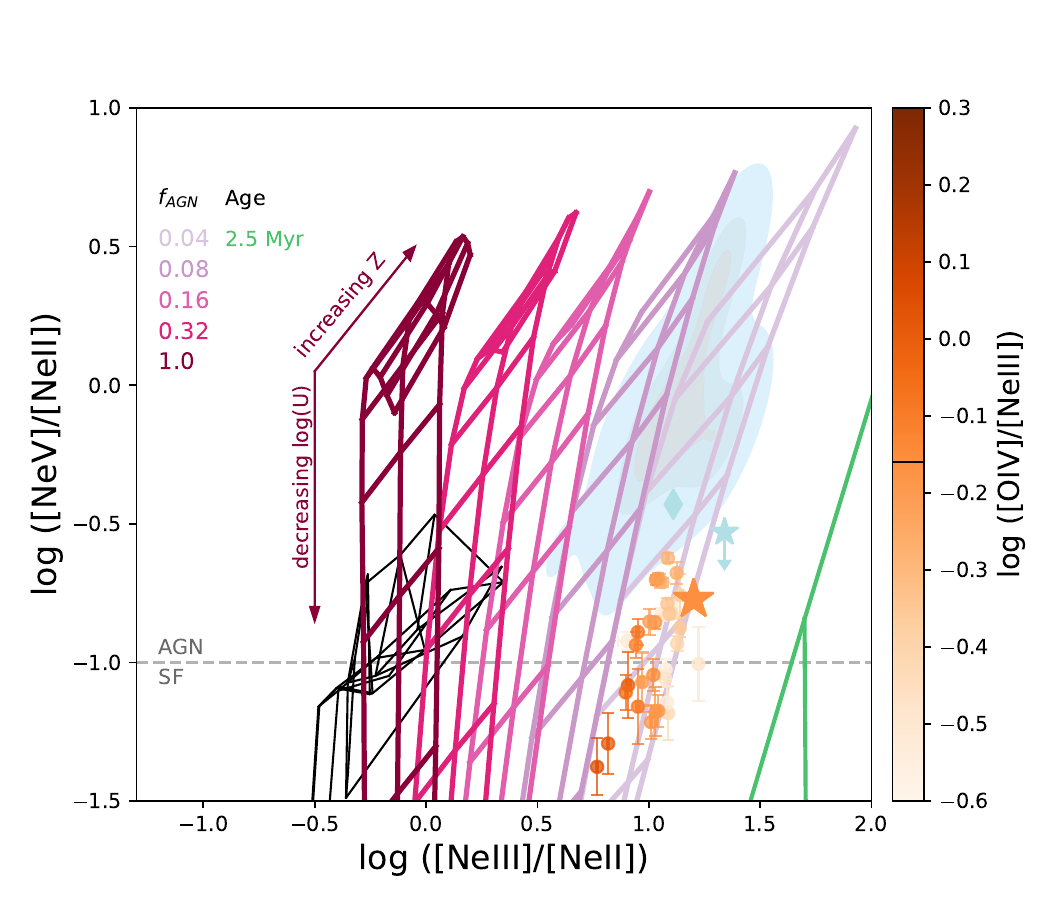}
    \includegraphics[width=0.5\textwidth]{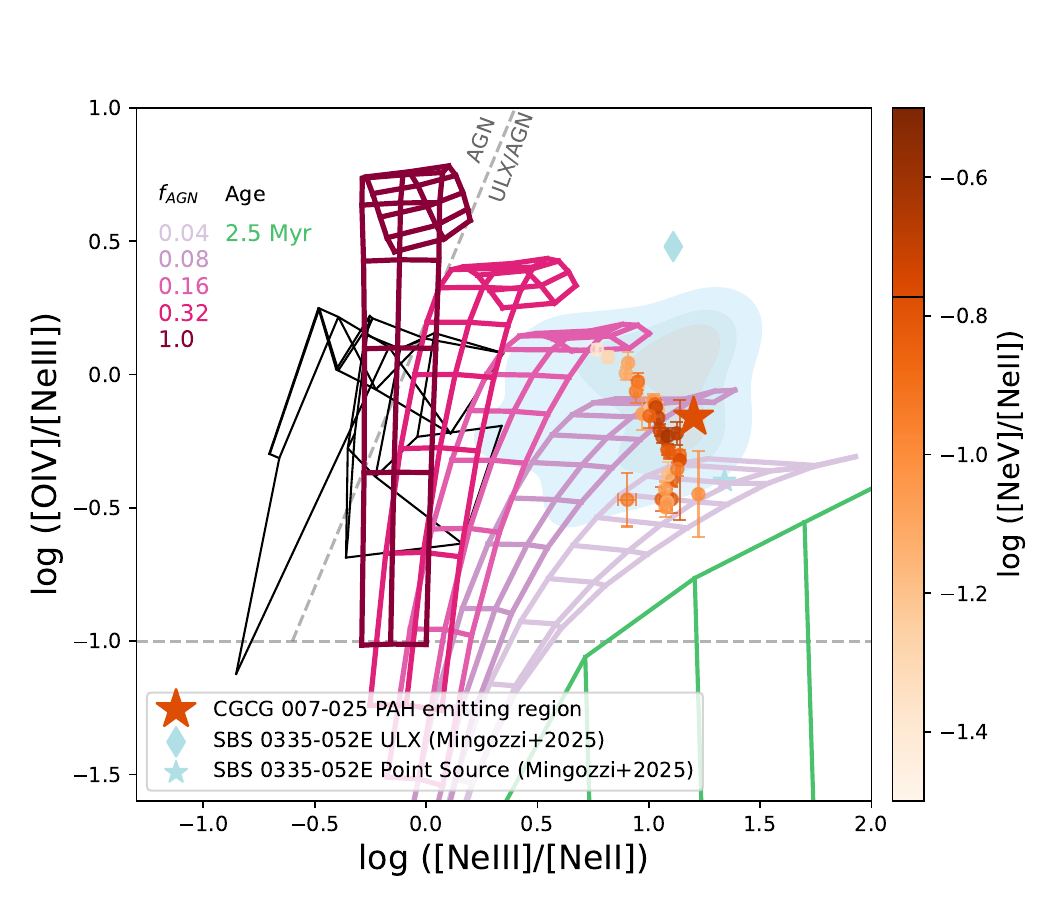}
    \caption{Excitation diagrams \nev/\neii~(\textit{top panel}) and \oiv/\neiii~(\textit{bottom panel}) vs. \neiii/\neii. Grids color-coded in purple-pink-white show the IMBH models from \citet{richardson2025arXiv250507749R} as a function of the AGN fraction. In green, the grid for a 2.5 Myr SSP from \citet{martinezparedes2023MNRAS.525.2916M}, and in black the shock models from \citet{flury2025MNRAS.tmp.1541F}. The star corresponds to the integrated line ratios for CGCG 007-025 within the PAH aperture, while the individual spaxels are shown as dots. Literature points for SBS 0335-052E are shown in light blue \citep[from][]{mingozzi2025ApJ...985..253M}: diamond for the ULX, star for the point-like source, and contours for the extended emission. The dashed gray in the top panel is the demarcation line from \citet{inami2013ApJ...777..156I}, while the dashed gray lines in the bottom panel are the demarcation lines from \citet{richardson2025arXiv250507749R}. The line ratios of CGCG 007-025 overlap with the 4\%-8\% AGN fraction grids.}
    \label{fig:bpts}
\end{figure}

We show in Figure \ref{fig:PAH_Nes} the 6.2/11.3 PAH ratio against \neiii/\neii, which is tracing the level of ionization. Local HII galaxies are concentrated in the upper left corner of the plot, with LIRGs (Luminous Infrared Galaxies) displaying similar \neiii/\neii~ratios but slightly lower 6.2/11.3 ratios. Seyfert galaxies move towards the center of the plot, showing an increase in the level of ionization and a further decrease in the 6.2/11.3 ratio. Blue Compact Dwarfs (BCDs), on the other hand, cover a wider PAH ratio range while showing even higher ionization. 
CGCG~007-025 (orange arrow in the lower right corner) stands out as an outlier in this diagram, showing one of the largest \neiii/\neii~ratio together with a deficit on the 6.2/11.3 PAH ratio. Other BCDs, like IIZw40 (${\rm Z \sim 0.25~Z_\odot}$; \citealt{lai2025arXiv250904662L}), display a larger \neiii/\neii~ratio, but their PAH ratios, and metallicity, are not as low as in the case of CGCG~007-025. 
The location of CGCG~007-025 in the diagram may indicate the presence of a highly ionizing source, as its PAH strengths are consistent with those typically observed in AGN-dominated systems. Indeed, its PAH and mid-IR line properties closely resemble those observed in the nucleus of the Seyfert~2 galaxy NGC~7319 and in radio galaxies with strong jets, which show only a marginal detection of the 11.3~$\mu$m PAH feature \citep{pereiresantaella2022A&A...665L..11P,garciabernete2022A&A...666L...5G,zakamska2016MNRAS.455.4191Z}. Similarly, the \neiii/\neii~and PAH~6.2/11.3 ratios of the AGN NGC~5728 \citep{garciabernete2024A&A...691A.162G} are consistent with those measured in this galaxy.

In addition to a significantly large \neiii/\neii~ratio, the integrated spectrum of CGCG~007-025 also displays high-ionization mid-IR lines such as \oiv\ (I.P.~$>$50 eV) and \nev\ (I.P.~$>$97 eV), which are hard to explain in pure star-forming environments \citep[e.g.,][]{lutz1998A&A...333L..75L,schaerer1999A&A...345L..17S}. 
Nevertheless, \oiv\ emission is commonly observed in BCDs \citep[e.g.,][]{dale2009ApJ...693.1821D,bernardsalas2009ApJS..184..230B}, and recent MIRI/MRS observations have revealed extended \nev\ in some extreme low-metallicity systems, such as SBS~0335–052E and I~Zw~18 (\citealt{mingozzi2025ApJ...985..253M,hunt2025arXiv250809251H}; see also e.g., \citealt{izotov2012MNRAS.427.1229I,izotov2021MNRAS.508.2556I} for detections on the optical). The origin of such high-ionization emission in low-metallicity environments remains puzzling and not yet fully understood \citep{berg2021ApJ...922..170B,olivier2022ApJ...938...16O}. 
We further explore this in Figure \ref{fig:bpts}, which displays the \nev/\neii\, vs \neiii/\neii~(top panel) and \oiv/\neiii\, vs \neiii/\neii~(bottom panel) excitation diagnostic diagrams. CGCG~007-025 PAH emitting region and emission line ratios at spaxel level are shown as a star and colored dot symbols, respectively, and color-coded in accordance to the complementary line ratio of each panel (the value of the star is marked in the colorbar). 
Overlaid in green are the single stellar population (SSP) models from \citet{martinezparedes2023MNRAS.525.2916M} for an age\footnote{This is the only age grid that appears within the displayed \nev/\neii, \neiii/\neii~and \oiv/\neiii~ranges and is the closest to the estimated age for CGCG 007-025 of 5 Myr from \citet{valleespinosa2023MNRAS.522.2089D}} of 2.5~Myr, with a top-heavy IMF and ${\rm M_{up} = 300 M_\odot}$ and metallicity ranging from ${\rm 12+log(O/H) = 6.94 - 7.94~(Z=0.02 - 0.2)}$. Interestingly, in both panels, the data points are far away from the locus of the pure SF models.

For the other two recently discovered \nev-emitters - SBS~0335–052E (light blue markers and contours in Figure~\ref{fig:bpts}) and I~Zw~18 - the only models capable of reproducing the observed line ratios require the presence of an additional source, such as an intermediate-mass black hole \citep[IMBH;][]{mingozzi2025ApJ...985..253M,hunt2025arXiv250809251H}.
Thus, we included in Figure \ref{fig:bpts} the IMBH grids from \citet{richardson2025arXiv250507749R} with ${\rm M_{BH}= 10^5 M_\odot}$, an age of 5 Myr, limited to a metallicity range of ${\rm 12+log(O/H) = 7.34 - 7.94~(Z=0.05 - 0.2)}$, and color-coded by the AGN fraction. 
Consistently with \citet{mingozzi2025ApJ...985..253M}, the line ratios of CGCG~007-025 could be explained with a modest 4-8\% of AGN fraction. 
Another possible source of ionization capable of generating \nev~could be shocks. We have included in Figure \ref{fig:bpts}, in black, the shock+precursor models from \citet{flury2025MNRAS.tmp.1541F} with 10\%~${\rm Z_\odot}$. 
However, these models fail to reproduce, simultaneously, the high and low ionization species seen in CGCG~007-025.

We note, however, that exotic line ratios are not enough to probe the existence of an IMBH in this system, and additional observables should be tested. For example, the presence of broad components in the permitted lines, not detected in other forbidden lines, or the presence of X-ray emission, are characteristic of AGNs. In this vein, \citet{senchyna2020MNRAS.494..941S} reported a prominent X-ray source located within the brightest SF region. Later, \citet{senchyna2022ApJ...930..105S} demonstrated that photoionization by extremely metal-poor stars, coupled with high-mass X-ray binaries or ultra luminous X-ray sources (ULXs), cannot reproduce the observed He II/H$\beta$ in this source. Moreover, in \citet{valleespinosa2023MNRAS.522.2089D} we measured a broad component in the H$\alpha$ line profile with velocity dispersion typical of AGN broad-line regions ($\sigma\sim$1000 km/s), although the lack of other evidences of AGN activity at that time precluded the determination of its origin. 
Overall, we would like to remark this is the first time that the emission line ratios of this galaxy fall outside the SF locus in diagnostic diagrams \citep[using e.g., UV and optical line ratios;][]{mingozzi2022ApJ...939..110M,valleespinosa2023MNRAS.522.2089D}. 
It remains possible that current stellar population models underestimate the production of high-energy ionizing photons, particularly in low-metallicity environments \citep[e.g.,][]{sellmaier1996A&A...305L..37S, stasinska2015A&A...576A..83S,fernandez-ontiveros2016ApJS..226...19F}. The present dataset therefore offers a valuable benchmark for guiding and constraining the next generation of star-formation models aimed at capturing such extreme conditions.

Complementing these diagnostics, \citet{garciabernete2024A&A...691A.162G} recently studied the relation between the PAH feature at 6.2 \mm~and the \hm~S(1) line to establish alternative ways to determine the presence of an AGN in the absence of high IP lines. The authors suggest that ratios of 6.2/S(1) larger than 12 are dominated by SF processes, whereas ratios below this value reflect an increase on the \hm~due to the heating of shocks/AGN. Stronger radiation fields would further destroy the PAHs, reducing the ratio below 5. For CGCG 007-025, we report a upper limit of ${\rm PAH_{6.2~\mu m}/H_2~S(1) < 0.71}$, which would fall in the AGN demarcation. Nevertheless, this trend has been established in (super-)solar metallicity environments as well as in confirmed AGN host galaxies.
At lower metallicities, it is still to be determined whether this ratio can be attributed to the harder ionizing spectrum (which would destroy the PAH complexes and increase the \hm~S(1) flux, simultaneously), or whether they are shaped simply because of the lack of PAH molecules in the first place.

It is worth noting that the PAH emission peaks at the location of high-IP lines (\oiv~and \nev), with no significant PAH detections in regions outside of these high-IP boundaries. In the scenario of PAH destruction in the presence of a highly ionizing field, this spatial anti-correlation seems contradictory: smaller PAHs should survive preferentially in lower-ionization environments. An alternative to PAH destruction is in-situ formation, where the source producing \oiv~and \nev~emission could also be the origin of PAH molecules. For example, the formation of PAH has previously been seen in the dusty winds of massive and/or Wolf-Rayet (WR) stars \citep{lau2022NatAs...6.1308L,taniguchi2025arXiv250901026T}. Although WR stars seem capable of producing \nev~emission \citep[see][]{tarantino2024ApJ...969..101T,hunt2025arXiv250809251H}, current photoionization models of massive stars cannot reproduce the mid-IR line ratios seen in CGCG~007-025. Further theoretical and observational efforts are thus required to understand the production of high-ionization photons and the formation of dust and complex molecules in low-metallicity starburst environments.

%
%

\section{Conclusions} \label{sec:conclusions}

In this letter, we present the first clear and spatially resolved PAH detection at ${\rm 10\%Z_\odot}$ in the dwarf galaxy CGCG 007-025 using the unique capabilities of JWST MIRI/MRS. Our findings are summarized below:

\begin{itemize}
    \item The 11.3 \mm~PAH feature is significantly detected at spaxel level (left panel of Figure \ref{fig:pahMapandSpectrum}). We explore the extension and location of this feature. The emission is spatially confined to a region of $\sim50$ pc, offset from the brightest clumps on the galaxy.
    \item Using \textsc{pahfit}, we characterize the continuum and spectral features (nebular lines and PAHs) of the integrated spectrum of the PAH emitting region (right panel of Figure \ref{fig:pahMapandSpectrum}). We confirm the presence of the 11.3 \mm~PAH feature with a flux of ${\rm F_{11.3 ~\mu m ~PAH} = (5.1 \pm 0.4) \times10^{-16}~erg/s/cm^2}$. 
    \item We find evidence of another PAH feature at 12.7 \mm, but the usually bright features at 6.2, 7.7 or 8.6 \mm~ are absent in CGCG 007-025.
    \item The detection of the 11.3 and 12.7 \mm~feature may indicate a PAH population dominated by larger, neutral molecules. Moreover, the 6.2/11.3 ratio, in conjunction with a large \neiii/\neii~ratio (Figure \ref{fig:PAH_Nes}), hint to the presence of a strong ionizing source, with \neiii/\neii~ratios compatible with those seen in harsh environments like AGNs or other BCDs.
    \item The presence of \nev~and \oiv~in the region is hard to explain using pure SF models. Instead, the line ratios are compatible with a mix of SF+AGN photoionization, with a moderate AGN fraction no larger than 8\% (Figure \ref{fig:bpts}). However, it remains possible that stellar models fail to produce enough high-ionizing photons.
\end{itemize}

To conclude, this study demonstrates that JWST/MIRI data is capable of not only detecting but also spatially resolving faint PAH emission in metal-poor environments. As a result, this instrument provides new insight into the effectiveness of mid-IR diagnostics and the nature of ionizing sources in systems previously considered to be purely star-forming systems, offering valuable guidance for understanding the physical conditions of dust and gas in metal-poor environments typical of the high-z universe.

%
%

\begin{acknowledgments}
The authors thank the anonymous referee for providing useful comments, which have certainly improved the quality of this letter. We also thank the MIRI/MRS support team for their guidance with the developer version of the pipeline. This work is based on observations made with the NASA/ESA/CSA James Webb Space Telescope. The data were obtained from the Mikulski Archive for Space Telescopes at the Space Telescope Science Institute, which is operated by the Association of Universities for Research in Astronomy, Inc., under NASA contract NAS 5-03127 for JWST. These observations are associated with program JWST-4278. 
MGVE is grateful for the support of this program, provided by NASA through a grant from the Space Telescope Science Institute.
The {\it JWST/MIRI-MRS} data used in this paper can be found in MAST at \dataset[doi: 10.17909/s0d0-jn32]{https://doi.org/10.17909/s0d0-jn32}.
MM, BLJ and SH are thankful for support from the European Space Agency (ESA).
MJH is supported by the Swedish Research Council (Vetenskapsr{\aa}det), and is fellow of the Knut \& Alice Wallenberg Foundation. 
RA acknowledges the support of project PID2023-147386NB-I00 and the Severo Ochoa grant CEX2021-001131-S funded by MCIN/AEI/10.13039/50110001103. LV acknowledges support from the INAF Minigrant “RISE: Resolving the ISM and Star formation in the Epoch of Reionization” (Ob. Fu. 1.05.24.07.01). 
This research has used the HSLA database, developed and maintained at STScI, Baltimore, USA. 
MGVE is grateful to Alberto Saldana-Lopez, Elizabeth Tarantino and Julia Roman-Duval for inspiring conversations and advice.

\end{acknowledgments}

\vspace{5mm}
\facilities{JWST(MIRI/MRS), HST(WFC3), }


\software{astropy \citep{2013A&A...558A..33A,2018AJ....156..123A,2022ApJ...935..167A},  
          }



\appendix

\section{Data reduction} \label{sec:appendixData}

We retrieved the mid-IR data of CGCG 007-025 from MAST. 
The galaxy was observed with JWST/MIRI MRS using the three MRS grating settings (SHORT, MEDIUM and LONG) and the FAST readout mode, giving full coverage in the wavelength range ${\rm \lambda \sim 5-28~\mu m}$. A single pointing was taken, centered in the main starbursting region of the galaxy. Given the JWST pointing accuracy of $\sim0.1$", target acquisition was not required. A 4-point dither pattern was implemented to ensure optimal sampling across the FoVs and to facilitate the identification and removal of detector artifacts. Each pointing was observed with 50 groups per integration, with a total exposure time of 1443 seconds. To enable accurate background subtraction, dedicated background observations were conducted using half the number of dithers. The background and science observations were placed in an uninterruptible sequence. 

The data were reduced using the JWST data reduction pipeline version \textit{1.17.1.dev} with the CRDS 1338 context. Briefly, in Stage~1 corrections at detector level are applied to produce data in count rate units. These include corrections for dark current, masking of detector cosmetics, and removal of strong jumps introduced by the impact of cosmic rays. In Stage~2, data is flat-field corrected, and wavelength and flux calibrated. A first inspection of the count rate images revealed intense cosmic ray events all across the detector which are difficult to mask. At the time of reducing the data, the developer version of the pipeline was the only release which included the \textit{straylight} correction step, a crucial step for correcting these intense cosmic ray events. Background subtraction is performed in Stage~3 using a master sky background created with the dedicated sky observations. We opted for this option rather than a pixel-by-pixel correction, since the later introduces more noise in the final data products. Stage~3 also combines the individual exposures into a single one and reconstructs the 3D datacubes per band. We also enabled a residual fringe correction in Stage~2 to account for any residual fringing not corrected during Stage~1. For extended sources, the simple fringe-flat correction applied in Stage~1 is usually enough to reduce its impact on the data products. However, the galaxy displays a PSF-like structure, due to an unresolved compact emission, making the fringe patterns vary systematically across the PSF and requiring further corrections to create the final extended mid-IR emission datacubes. We chose the \textit{emsm} algorithm as a first step to minimize the wiggle patterns of the PSF. The additional PSF-like emission modeling and subtraction was done following the \citet{mingozzi2025ApJ...985..253M} procedure. Besides our best efforts to correct from fringing and PSF undersampling, some spaxels situated at the peak of the mid-IR emission are still affected by these wiggling patterns. We also verified that the astrometry of the MIRI/MRS datacubes matched the World Coordinate System (WCS) of the available HST data of this galaxy, as well as the WCS of the MUSE dataset from which we obtained the H$\alpha$ contours.

\section{Absence of typical PAH features in CGCG 007-025} \label{sec:appendixPAHs}

\begin{figure*}
    \centering
    \includegraphics[width=0.45\textwidth]{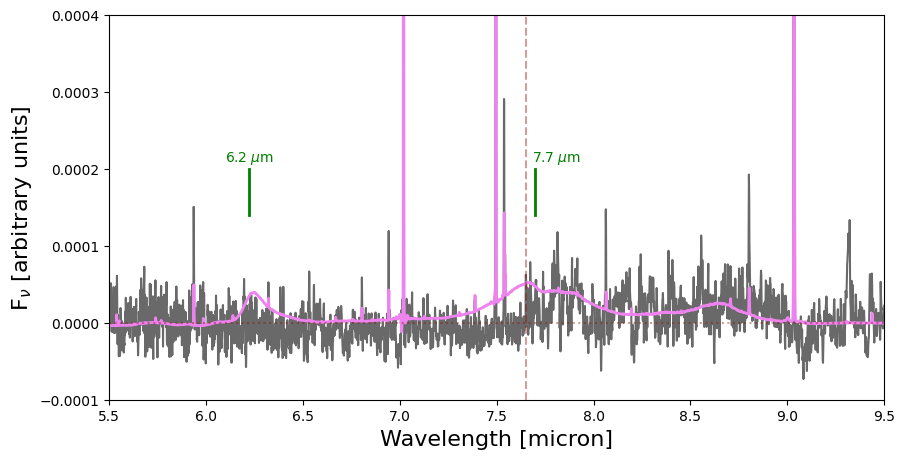}
    \includegraphics[width=0.45\textwidth]{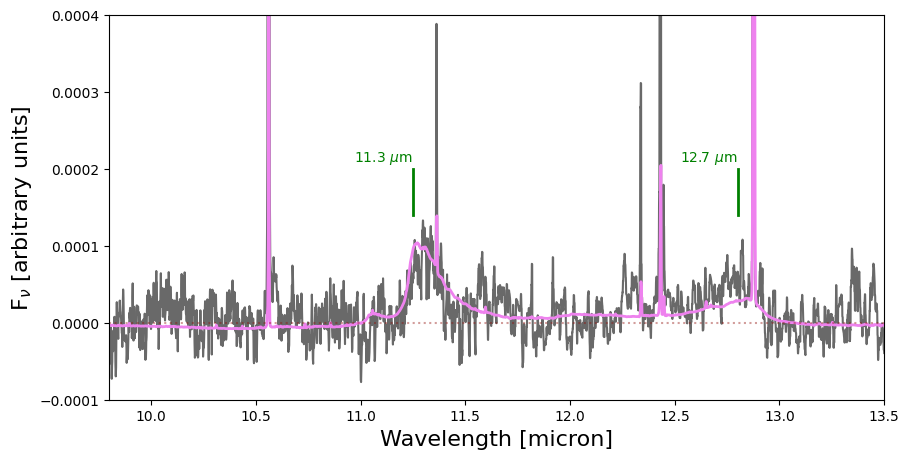}
    \caption{Comparison between the integrated spectrum of CGCG 007-025 (gray line) with the PDRs4All model of HII regions (pink line). The spectra have been continuum subtracted, and the PDRs4All model has been re-scaled to match the flux on the 11.3$\mu m$ PAH feature.}
    \label{fig:appendixpahs}
\end{figure*}

Despite of the bump at 11.3 $\mu$m, the integrated spectrum of CGCG 007-025 does not display clear evidence of other PAH features. In Figure \ref{fig:appendixpahs}, we compare the integrated spectrum of CGCG 007-025 (gray line) with the PDRs4All template for HII regions \citep[pink line,][]{chown2024A&A...685A..75C}. Both spectra have been continuum subtracted, and the template is also re-scale to match the flux of the 11.3 $\mu$m feature. The spectrum of CGCG 007-025 shows no excess of emission at 6.2 and 7.7 $\mu$m (left panel), indicating no evidence of these features in the aperture. However, the spectrum displays a slight excess of emission at 12.7 $\mu$m, matching the aforementioned template (right panel). We measure the flux associated with the 12.7 \mm~feature on the continuum subtracted spectrum using a Drude profile. The integrated flux, corrected for attenuation, can be found in Table \ref{tab:fluxes}. Following \citet{lai2025arXiv250904662L}, we derive upper limits on the undetected PAHs by estimating the $3\sigma$ integrated fluxes of the 6.2, 7.7 and 8.6 \mm~bands. We define $\sigma$ as the product of the width of the PAH profile and rms of the residuals from PAHFIT in the wavelength range 9.68–10.43 \mm, where no PAH or emission lines are observed. The width of the PAH bands are defined as the interval spanning one FWHM, centered on the central wavelength \citep[][]{smith2007ApJ...656..770S}. For the PAH complexes 6.2 and 7.7, the wavelength range expands from half a FWHM below the bluest subcomponent to half a FWHM above the reddest subcomponent. The upper limits on the PAH bands can be found on Table \ref{tab:fluxes}.

\section{Additional material}\label{sec:appendixSFR}

In Table \ref{tab:fluxes} we compile the well-detected list of emission lines included in \textsc{pahfit} for the modeling of the PAH emitting region, as well as the unattenuated emission line measurements from \textsc{pahfit}. We also included the flux of the 11.3 and 12.7 $\mu$m PAH features and the upper limits on the 6.2, 7.7 and 8.6 $\mu$m PAH features.

\begin{table}
    \centering
    \begin{tabular}{l ccc ||ccc||cc} 
        \multicolumn{4}{c||}{Fine structure lines} & \multicolumn{3}{c||}{\hm~lines} & \multicolumn{2}{c}{PAH} \\ 
        Line name & IP & $\lambda_0$ & Flux & Line name & $\lambda_0$ & Flux & Feature & Flux \\
         & [eV] & [\mm] & [$\rm 10^{-16}~erg/s/cm^2$]& & [\mm] & [$\rm 10^{-16}~erg/s/cm^2$]& & [$\rm 10^{-16}~erg/s/cm^2$]\\ \tableline

         \multirow{2}{*}{\textbf{\feii}}         & \multirow{2}{*}{7.9}   & \multirow{2}{*}{5.34}  & 0.4 ~ $\pm$ ~0.3 & \multirow{2}{*}{S(1)} & \multirow{2}{*}{17.035} & 0.80 $\pm$ 0.03 & \multirow{2}{*}{6.2 \mm} & $\leq 0.48 ~(3\sigma)$ \\
          &  &  &  (0.5~ $\pm$ ~0.3)&  &  & (1.00 $\pm$ 0.03) &  & ($\leq 0.57 ~(3\sigma)$) \\ 
         \multirow{2}{*}{\arii}                  & \multirow{2}{*}{15.76} & \multirow{2}{*}{6.99}  & 0.29~ $\pm$ ~0.08 & & & & \multirow{2}{*}{7.7 \mm} & $\leq 0.54 ~(3\sigma)$ \\
          &  &  & (0.3~ $\pm$ ~0.1)& & & &  & ($\leq 0.62 ~(3\sigma)$) \\ 
         \multirow{2}{*}{\ariii}                 & \multirow{2}{*}{27.63} & \multirow{2}{*}{8.99}  & 3.86~ $\pm$ ~0.08 & & & & \multirow{2}{*}{8.6 \mm} & $\leq 0.39 ~(3\sigma)$ \\
         & & & (5.8~ $\pm$ ~0.1)& & & & & ($\leq 0.50 ~(3\sigma)$) \\ 
         \multirow{2}{*}{\siv}                   & \multirow{2}{*}{34.83} & \multirow{2}{*}{10.51} & 31.2~ $\pm$ ~0.4  & & & & \multirow{2}{*}{11.3 \mm} & 3.9 $\pm$ 0.4 \\
         & & & (49.0~ $\pm$ ~0.6)& & & & & (5.1 $\pm$ 0.4) \\ 
         \multirow{2}{*}{\neii$^\star$ }         & \multirow{2}{*}{21.56} & \multirow{2}{*}{12.81} & 2.22~ $\pm$ ~0.03 &  & & & \multirow{2}{*}{12.7 \mm} & 7.0 $\pm$ 0.4 \\
         & & & (2.38~ $\pm$ ~0.03)& & & &  & (7.5 $\pm$ 0.4) \\
         \multirow{2}{*}{\textbf{\nev}$^\star$}  & \multirow{2}{*}{97.12} & \multirow{2}{*}{14.32} & 0.38~ $\pm$ ~0.02 & & & & & \\
         & & & (0.41~ $\pm$ ~0.02)& & & & & \\ 
         \multirow{2}{*}{\neiii$^\star$}         & \multirow{2}{*}{40.96} & \multirow{2}{*}{15.55} & 35.3~ $\pm$ ~0.4  &  & & & & \\
         & & & (40.3~ $\pm$ ~0.5)& & & & & \\ 
         \multirow{2}{*}{\siii}                  & \multirow{2}{*}{23.34} & \multirow{2}{*}{18.71} & 43.4~ $\pm$ ~0.4  & & & & & \\
         & & & (55.1~ $\pm$ ~0.5)& & & & & \\ 
         \multirow{2}{*}{\oiv$^\star$}           & \multirow{2}{*}{54.95} & \multirow{2}{*}{25.89} & 24.4~ $\pm$ ~0.4  & & & & & \\
         & & & (25.4~ $\pm$ ~0.4)& & & & & \\ 
         \multirow{2}{*}{\feii}                  & \multirow{2}{*}{7.9}   & \multirow{2}{*}{25.99} & 1.2 ~ $\pm$ ~0.3  &  & & & & \\
         & & & (1.2~ $\pm$ ~0.3)& & & & & \\ 

    \end{tabular}
    \caption{Fluxes of the different emission lines and PAH features as modeled by \textsc{pahfit} for the PAH integrated spectrum. Fluxes reported in brackets have been corrected for intrinsic attenuation. Transitions in bold were not present in the original \textsc{pahfit} version \citep{smith2007ApJ...656..770S}. Spectral lines marked with $^\star$ are the focus of the extended emission modeling (see Sec. \ref{sec:methext}).}
    \label{tab:fluxes}
\end{table}

\bibliography{coreTex}{}
\bibliographystyle{aasjournal}

\end{document}